\documentstyle[12pt]{article}
\font\frakbig=eufm10 scaled 1728
\begin{document}
\def\eea{\end{eqnarray}}
\def\nn{\nonumber}
\def\ZZ{\Bbb{Z}}
\def\bea{\begin{eqnarray}}
\def\CC{\Bbb{C}}
\def\dis#1{\displaystyle{#1}}
\def\stackreb#1#2{\mathrel{\mathop{#2}\limits_{#1}}}
\def\res#1{\stackreb{#1}{\rm res}}
\def\ep{\varepsilon}
\def\Uqap{U_q({\frak sl}^\prime_2)}
\def\Uqanp{U_q({\frak sl}^\prime_n)}
\def\Uqa{U_q(\widehat{\frak{sl}}_2)}
\def\Uqan{U_q(\widehat{\frak{sl}}_n)}
\def\Uqn{U_q(\frak{sl}_n)}
\def\slaff{\frak{sl}^\prime_2}
\def\aff{\widehat{\frak{sl}}_2}
\def\da{\partial_\alpha}
\def\db#1{\partial_{\beta^{(#1)}}}
\def\p#1{\psi^{(#1)}}
\def\ps#1{\psi^{(#1)*}}
\def\vf#1{\varphi^{(#1)}}
\def\ee{{\rm e}}
\def\ip#1{(#1)_\infty}
\def\ipp#1#2{(#1;#2)_\infty}
\catcode`\@=11
\ifcase\@ptsize
 \font\tenmsy=msbm10
 \font\sevenmsy=msbm7
 \font\fivemsy=msbm5
 \font\teneu=eufm10
 \font\seveneu=eufm7
 \font\fiveeu=eufm5
\or
 \font\tenmsy=msbm10 scaled \magstephalf
 \font\sevenmsy=msbm8
 \font\fivemsy=msbm6
 \font\teneu=eufm10 scaled \magstephalf
 \font\seveneu=eufm8
 \font\fiveeu=eufm6
\or
 \font\tenmsy=msbm10 scaled \magstep1
 \font\sevenmsy=msbm8
 \font\fivemsy=msbm6
\font\teneu=eufm10   scaled \magstep1
\font\seveneu=eufm8
\font\fiveeu=eufm6
\fi
\newfam\msyfam
\textfont\msyfam=\tenmsy  \scriptfont\msyfam=\sevenmsy
  \scriptscriptfont\msyfam=\fivemsy
\def\Bbb{\ifmmode\let\next\Bbb@\else
 \def\next{\errmessage{Use \string\Bbb\space only in math mode}}\fi\next}
\def\Bbb@#1{{\Bbb@@{#1}}}
\def\Bbb@@#1{\fam\msyfam#1}
\newfam\eufam
\textfont\eufam=\teneu  \scriptfont\eufam=\seveneu
  \scriptscriptfont\eufam=\fiveeu
\def\frak{\ifmmode\let\next\frak@\else
 \def\next{\errmessage{Use \string\frak\space only in math mode}}\fi\next}
\def\frak@#1{{\frak@@{#1}}}
\def\frak@@#1{\fam\eufam#1}
\catcode`\@=12
\makeatletter
\newdimen\normalarrayskip              
\newdimen\minarrayskip                 
\normalarrayskip\baselineskip
\minarrayskip\jot
\newif\ifold             \oldtrue            \def\new{\oldfalse}
\def\arraymode{\ifold\relax\else\displaystyle\fi} 
\def\eqnumphantom{\phantom{(\theequation)}}     
\def\@arrayskip{\ifold\baselineskip\z@\lineskip\z@
     \else
     \baselineskip\minarrayskip\lineskip2\minarrayskip\fi}
\def\@arrayclassz{\ifcase \@lastchclass \@acolampacol \or
\@ampacol \or \or \or \@addamp \or
   \@acolampacol \or \@firstampfalse \@acol \fi
\edef\@preamble{\@preamble
  \ifcase \@chnum
     \hfil$\relax\arraymode\@sharp$\hfil
     \or $\relax\arraymode\@sharp$\hfil
     \or \hfil$\relax\arraymode\@sharp$\fi}}
\def\@array[#1]#2{\setbox\@arstrutbox=\hbox{\vrule
     height\arraystretch \ht\strutbox
     depth\arraystretch \dp\strutbox
     width\z@}\@mkpream{#2}\edef\@preamble{\halign \noexpand\@halignto
\bgroup \tabskip\z@ \@arstrut \@preamble \tabskip\z@ \cr}%
\let\@startpbox\@@startpbox \let\@endpbox\@@endpbox
  \if #1t\vtop \else \if#1b\vbox \else \vcenter \fi\fi
  \bgroup \let\par\relax
  \let\@sharp##\let\protect\relax
  \@arrayskip\@preamble}
%
%
%
%
\def\eqnarray{\stepcounter{equation}%
              \let\@currentlabel=\theequation
              \global\@eqnswtrue
              \global\@eqcnt\z@
              \tabskip\@centering
              \let\\=\@eqncr
              $$%
 \halign to \displaywidth\bgroup
    \eqnumphantom\@eqnsel\hskip\@centering
    $\displaystyle \tabskip\z@ {##}$%
    &\global\@eqcnt\@ne \hskip 2\arraycolsep
         $\displaystyle\arraymode{##}$\hfil
    &\global\@eqcnt\tw@ \hskip 2\arraycolsep
         $\displaystyle\tabskip\z@{##}$\hfil
         \tabskip\@centering
    &{##}\tabskip\z@\cr}
\makeatother
\begin{center}
\hfill DFTUZ/94/15\\
\hfill hep-th/9410021\\
\bigskip\bigskip
{\Large  On the Bosonization of $L$-Operators for\\
Quantum Affine Algebra $U_q(\hbox{\frakbig sl}_2)$}
\footnote{Talk given at the International Coference "Modern Problems
of Quantum Field Theory, Quantum Gravity and Strings", Alushta,
June 10--20, 1994}\\
\bigskip
\bigskip
{\large S. Pakuliak}\footnote{E-mail:
pakuliak@cc.unizar.es}
\footnote{On leave of absence from the ITP, Kiev 252143, Ukraine}\\
\bigskip
{\it Departamento de F\'\i sica Te\'orica\\
Facultad de Ciencias\\
Universidad de Zaragoza\\
50009 Zaragoza, Spain}\\
\bigskip
\end{center}
\begin{abstract}
Some relations between different objects associated with
 quantum affine algebras are reviewed.
It is shown that the Frenkel-Jing bosonization of a new realization
of quantum affine algebra $\Uqa$ as well as bosonization of $L$-operators
for this algebra
can be obtained from Zamolodchikov-Faddeev algebras defined by the
quantum $R$-matrix satisfying unitarity and crossing-symmetry
conditions.
\end{abstract}

\section{Introduction}
\setcounter{footnote}{0}
Recently there have been arised a great interest
to the free field
realization of the Zamolod\-chi\-kov-Faddeev (ZF)
algebras \cite{ZZ79} appearing in the context of quantum integrable
models \cite{DFJMN93,JMMN92}. These realizations are important
for calculations of the correlation function and form factors
in these models.  It was shown in \cite{DFJMN93,FR92} that,
if $R$-matrix
defining the ZF algebra is intertwiner of finite-dimensional
representations of the quantum affine algebra $\Uqa$, then one can
naturally realize the corresponding ZF algebras
in terms of operators that intertwine the infinite-dimensional
highest or lowest weight modules over $\Uqa$.

The algebra $\Uqa$ was introduced by Drinfeld and Jimbo
\cite{Dr87,Ji85} using deformed relations between Chevalley
generators. Lately in the framework of quantum inverse scattering
method or in
``the $RLL$ formalism'' this algebra was formulated in terms of $L$-operators
\cite{RS90}. Quite recently, it has been understood that these operators
 can be expressed via  operators satisfying
ZF algebras associated with the quantum
$R$-matrix acting in the tensor product of the finite-dimensional
representation spaces of $\Uqa$ \cite{Mi94}.

So called ``new realization of quantum affine algebras''
was introduced  in \cite{Dr88} which
appeared to be a quantum analog of the loop realization of affine
algebras. I.~Frenkel and J.~Ding have found an isomorphism
between the $R$-matrix formulation of $\Uqan$  and    the
new realization for this algebra \cite{FD93}.

Fixing level of $\Uqa$ to be unity, one we can easily obtain
the free field realization of all these objects starting
from Frenkel-Jing bo\-soni\-zation of level one $\Uqa$ \cite{FJ88}.
Using Hopf structure of  the algebra $\Uqa$    one can bosonize
the corresponding ZF algebras following \cite{JMMN92}.
Then, using Miki's formulas (\ref{25}), the bosonization of $L$-operators
can be immediately obtained.

The aim of this note is to show that it is possible to reconstruct
the non-abelian symmetry algebra of the model
starting from  the  corresponding
ZF algebras.
In this letter
we consider only the case of ZF algebras related to  the
XXZ model in anti-ferroelectric   regime on the infinite lattice.
The symmetry algebra for this model is the
quantum affine algebra $\Uqa$ with $-1<q<0$.
Let us formulate the main observation of this note.
\smallskip

{\sl A new realization of the quantum affine algebra $\Uqa$ at level one
follows from the free field realization of Zamolodchikov-Faddeev algebras
associated with $R$-matrix which satisfies
the conditions of unitarity and crossing-symmetry.}
\smallskip

Let us note that a similar approach to the bosonization of massive integrable
models in the field theory
was developed by S. Lukyanov in \cite{L93}. For
$SU_q(n)$ invariant spin chain on the finite lattice
the problem of reconstruction of the non-abelian
symmetry algebra from Yang-Baxter operators have been solved in
\cite{VG94}.

The paper is organized as follows. In Sect. 2 we will give a short review
of all objects that we mentioned  in Introduction. Sect. 3 is devoted
to the free field realization of ZF algebras. In conclusion
we will  list the problems, which the approach developed in this letter
might be also applied to.

\section{Algebra $U_q(\widehat{\hbox{\frakbig sl}}_2)$ in
Different Realizations}

We start with definition of $\Uqa$ following \cite{Ji85}.
Let $P$ and $P^*$ will be weight and dual weight lattices generated
respectively by $\{\Lambda_0,\Lambda_{1},\delta\}$  and
$\{h_0,h_{1},d\}$ with the canonical pairing
$$\langle \Lambda_i,h_j\rangle=\delta_{ij}, \quad
 \langle \Lambda_i,d\rangle=0,\quad \langle \delta,h_i\rangle=0,
\quad \langle\delta,d\rangle=1.
$$
Let $\alpha_0=2\Lambda_0-2\Lambda_{1}+\delta$ and
$\alpha_1=2\Lambda_1-2\Lambda_{0}$ be the simple roots.
Algebra $\Uqa$ is generated by the symbols
$t_i,t_i^{-1},e_i,f_i,q^d$ $i=0,1$ subjected to the following relations
$$
t_it_j=t_jt_i,\quad
t_ie_jt_i^{-1}=q^{\langle\alpha_j,h_i\rangle}e_j,\quad
t_if_jt_i^{-1}=q^{-\langle\alpha_j,h_i\rangle}f_j,
$$
$$
[e_i,f_j]=\delta_{ij}{t_i-t_i^{-1}\over q-q^{-1}},\quad
q^de_jq^{-d}=q^{\delta_{j0}}e_j,\quad
q^df_jq^{-d}=q^{-\delta_{j0}}f_j,
$$
$$
\sum_{k=0}^{3} (-)^k {[3]!\over[k]![3-k]!}e_i^ke_je_i^{3-k}=0,\quad
\sum_{k=0}^{3} (-)^k {[3]!\over[k]![3-k]!}f_i^kf_jf_i^{3-k}=0.
$$
Throughout the paper we will use the standard notation
$[n]=(q^n-q^{-n})/(q-q^{-1})$ and $[k]!=[1][2]\cdots[k]$.

Let $V=\CC\, v_++\CC\, v_-$ be a two-dimensional space and
$V_z=V\otimes\CC[z,z^{-1}]$ be its affini\-za\-tion \cite{KKMMNN92}.
Define the action
of the generators $t_i,t_i^{-1},e_i,f_i$  on the space $V_z$
as follows
$$
\pi_z(e_i)v_\ep z^n=\pi(e_i)v_\ep z^{n+\delta_{i0}},\quad
\pi_z(f_i)v_\ep z^n=\pi(f_i)v_\ep z^{n-\delta_{i0}},\quad
\pi_z(t_i)v_\ep z^n=\pi(t_i)v_\ep z^{n},
$$
where the action on the space $V$ is given
$$
\pi(e_0)=\pi(f_1)=
\left(
\begin{array}{cc}
0 &1 \\ 0 & 0
\end{array}\right),\quad
\pi(f_0)=\pi(e_1)=
\left(
\begin{array}{cc}
0 &0 \\ 1 & 0
\end{array}\right),\quad
\pi(t_0^{-1})=\pi(t_1)=
\left(
\begin{array}{cc}
q &0 \\ 0&q^{-1}
\end{array}\right)
$$
These formulas define the fundamental (vector) representation
of algebra $\Uqa$.

$\Uqa$ is a Hopf algebra.
Using the Hopf structure we can consider
the tensor products of fundamental representations and calculate the
intertwining operator ($R$-matrix) between these tensor products
$$
\overline
R_{12}(z_1/z_2) : V_{z_1} \otimes V_{z_2} \to  V_{z_2} \otimes V_{z_1}.
$$
With normalization $\overline R_{12}(z)v_+\otimes v_+ =  v_+\otimes v_+ $
$R$-matrix reads as follows
\bea\label{R-mat}
\overline{R}_{12}(z)&=&
\left(
\begin{array}{cccc}
1 &0 &0 &0 \\ 0& b(z) & c(z) &0 \\ 0 & \bar c(z) & b(z) &0 \\ 0&0 &0 & 1
\end{array}\right)\\
c(z)&=&\bar c(z)z={(1-q^2)z\over 1-q^2z}, \quad
b(z)={(1-z)q\over 1-q^2z},\nn
\eea

In what follows we
will define ZF algebras with $R$-matrix satisfying additional
conditions of unitarity
$$
R_{12}(z)R_{21}(z^{-1})=1
$$
and crossing-symmetry
$$
R_{12}(z^{-1})=(\sigma^x\otimes1)R_{12}(q^2z)(\sigma^x\otimes1),
$$
where
$$
\sigma^x=
\left(
\begin{array}{cc}
0&1 \\ 1&0
\end{array}\right).
$$
These conditions along with the requirement  that  $R$-matrix is an
analytic function for $q^2\leq |z|\leq q^{-2}$ uniquely determine
two normalization factors
$$
r^\pm(z)=\pm{1\over\sqrt{z}}{\ip{q^4z^{-1}}\ip{q^2z}\over
\ip{q^4z}\ip{q^2z^{-1}} },\quad
(a)_\infty=\prod_{n=0}^\infty (1-aq^{4n})
$$
in front of $\overline R_{12}(z)$,
so we can define two ZF algebras
\bea
\Phi(z_2)\Phi(z_1)&=&r^+(z)\overline R_{12}(z_1/z_2)\Phi(z_1)\Phi(z_2),
\nn\\
\Psi(z_1)\Psi(z_2)&=&r^-(z)\overline R_{12}(z_1/z_2)\Psi(z_2)\Psi(z_1).
\nn
\eea
In components these relations are
\bea
\Phi_{\ep_2}(z_2)\Phi_{\ep_1}(z_1)&=&r^+(z)\overline
R_{\ep_1\ep_2}^{\ep'_1\ep'_2}(z_1/z_2)\Phi_{\ep'_1}(z_1)\Phi_{\ep'_2}(z_2),
\label{typeI}\\
\Psi_{\ep_1}(z_1)\Psi_{\ep_2}(z_2)&=&r^-(z)\overline
R_{\ep_1\ep_2}^{\ep'_1\ep'_2}(z_1/z_2)\Psi_{\ep'_2}(z_2)\Psi_{\ep'_1}(z_1),
\label{typeII}
\eea
where indices $\ep_i$ take the values $\pm$ and we assume the summation
over repeated indices.
We will see in the next section
that the operators $\Phi$ and $\Psi$
will correspond to type I and type II operators in
terms of the paper \cite{DFJMN93}.

Let us formulate the algebra $\Uqa$ in $R$-matrix approach following
\cite{RS90}. Let
\begin{equation}
\tilde R(z)=\rho(z)\overline R(z)
\label{universal}
\end{equation}
be an
universal ${\cal R}$-matrix for the algebra $\Uqa$
evaluated on the tensor product of two
vector representations of $\Uqa$. Obviously it should coincide with
$R$-matrix (\ref{R-mat}) up to normalization factor. This normalization
factor  $\rho(z)$
can be calculated following \cite{FR92}  and is equal  to
$$
\rho(z)={1\over\sqrt{q}}{\ip{q^2z}^2\over
\ip{z}\ip{q^{4}z} } .
$$
Algebra $\Uqa$ at level one is generated by the coefficients of the
matrix series
\begin{equation}
L^{(\pm)}(z)  =  \sum_{n=0}^{\infty}   L^{(\pm)}_n z^{(\pm n)},
\label{L-operators}
\end{equation}
satisfying the relations
\begin{equation}
\tilde{R}_{12}(z_1/z_2)L_1^{(\pm)}(z_1)L_2^{(\pm)}(z_2)=
 L_2^{(\pm)}(z_2) L_1^{(\pm)}(z_1) \tilde{R}_{12}(z_1/z_2),
\label{26}
\end{equation}
\begin{equation}
\tilde{R}_{12}(z_1q/z_2)L_1^{(+)}(z_1)L_2^{(-)}(z_2)=
 L_2^{(-)}(z_2) L_1^{(+)}(z_1) \tilde{R}_{12}(z_1q^{-1}/z_2).
\label{28}
\end{equation}
In components these relations read as follows
$$
\tilde{R}_{\ep_1\ep_2}^{\ep'_1\ep'_2}(z_1/z_2)
L_{\ep'_1\nu_1}^{(\pm)}(z_1) L_{\ep'_2\nu_2}^{(\pm)}(z_2)=
L_{\ep_2\nu'_2}^{(\pm)}(z_2) L_{\ep_1\nu'_1}^{(\pm)}(z_1)
\tilde{R}_{\nu_1\nu_2}^{\nu'_1\nu'_2}(z_1/z_2) ,
$$
$$
\tilde{R}_{\ep_1\ep_2}^{\ep'_1\ep'_2}(z_1q/z_2)
L_{\ep'_1\nu_1}^{(+)}(z_1) L_{\ep'_2\nu_2}^{(-)}(z_2)=
L_{\ep_2\nu'_2}^{(-)}(z_2) L_{\ep_1\nu'_1}^{(+)}(z_1)
\tilde{R}_{\nu_1\nu_2}^{\nu'_1\nu'_2}(z_1q^{-1}/z_2).
$$
In fact, the relations
(\ref{26}) and (\ref{28}) define algebra $U_q(\widehat{\frak{gl}}_2)$.
In order to obtain $\Uqa$ we have to impose two more relations
that fix a quantum determinant of the operators $L^{\pm}(z)$
\footnote{By the quantum determinant of the operator
$\left(\begin{array}{cc}a(z)&b(z)\\c(z)&d(z)\end{array}\right)$
we  understand the following combination
$a(zq^{-2})d(z)-qb(zq^{-2})c(z)$.}
\begin{equation}
\hbox{\rm q-det}L^{(\pm)}
= -q. \label{q-det}
\end{equation}
The statement that quantum determinants $\hbox{\rm q-det}L^{(\pm)}$
belong to the center of
the algebra can be easily checked using the standard arguments of
quantum inverse scattering method and the fact that, after a proper
rescaling $R$-matrix
(\ref{universal}) in the point $z=q^{-2}$  has a simple form
$$
\left.{1\over \sqrt q} {(q^2)^2_\infty\over (q^4)^2_\infty}{1\over 1-zq^{2}}
\tilde{R}_{12}(z)\right|_{z=q^{-2}}=
\left(
\begin{array}{cccc}
0 &0 &0 &0 \\ 0& 1 & -q^{-1} &0 \\ 0 &  -q & 1 &0 \\ 0&0 &0 & 0
\end{array}\right).
$$


Our aim now is to relate the ZF algebras (\ref{typeI}) and
(\ref{typeII}) to algebra (\ref{26}), (\ref{28}). It can be done
following \cite{Mi94}. The bilinear combinations of the ZF
operators $\Phi_\varepsilon(z)$ and $\Psi_\varepsilon(z)$
($\ep,\nu=\pm$)
\bea
L^{(+)}_{\ep\nu} &=& \sqrt{z}g
\Psi_\nu(z)\Phi_\ep(zq)\nn\\
L^{(-)}_{\ep\nu} &=& \sqrt{z}g
\Phi_\ep(z)\Psi_\nu(zq)\label{25} \\
g&=&(q^{2})_\infty/(q^{4})_\infty   \nn
\eea
will satisfy the relations  (\ref{26}) and (\ref{28}) provided the operators
$\Psi_{\ep}(z)$ and $\Phi_{\nu}(z)$ commute with the scalar
function $\tau(z)$
\begin{equation}
\Psi_{\nu}(z_1)\Phi_{\ep}(z_2) =\tau\left(z_1/ z_2\right)
\Phi_{\ep}(z_2)\Psi_{\nu}(z_1)
\label{comm}
\end{equation}
which is defined by
$$
\tau^2(z)={r^+(zq)\over r^-(zq^{-1})}{\rho(zq^{-1})\over \rho(z q)}=
{1\over{z}}{\ip{q^3z^{-1}}^2\ip{qz}^2\over
\ip{q^3z}^2\ip{qz^{-1}}^2 }
$$
and satisfies the relation
$$
\tau(zq)\tau(zq^{-1})=-1.
$$


Let us describe  now
a relation between algebra (\ref{26}), (\ref{28})
and the new realization of quantum affine algebra \cite{Dr88}
following \cite{FD93}. Let
$$
L^\pm(z)=
\left(\begin{array}{cc} 1& e^\pm(z)\\ 0&1 \end{array}\right)
\left(\begin{array}{cc} k^\pm_1(z)&0\\ 0&k^\pm_2(z) \end{array}\right)
\left(\begin{array}{cc} 1& 0\\ f^\pm(z)&1 \end{array}\right)
$$
be the decomposition of $L$-operators. Then the operators
$$
\psi(z)=k_2^-(zq)k_1^-(zq)^{-1},\quad   \phi(z)=k_2^+(zq)k_1^+(zq)^{-1}
$$
$$
x^+(z)={e^+(zq^{-1/2})-e^-(zq^{1/2})\over q-q^{-1}},\quad
x^-(z)={f^+(zq^{1/2})-f^-(zq^{-1/2})\over q-q^{-1}}
$$
satisfy the commutation relations of the algebra $\Uqa$ at level 1
in the new realization
\bea
\psi(z)\phi(w)&=&{(z-wq^3)(z-wq^{-3})\over(z-wq)(z-wq^{-1})}\phi(w)
\psi(z)\label{21a}\\
\psi(z)x^\pm(w)&=&q^{\pm2}
{z-q^{\mp5/2}w\over z-q^{\pm3/2}w}x^\pm(w)\psi(z) \label{21b1}\\
\phi(z)x^\pm(w)&=&q^{\pm2}
{z-q^{\mp3/2}w\over z-q^{\pm5/2}w}x^\pm(w)\phi(z)   \label{21c1}\\
x^\pm(z)x^\pm(w)&=& {(zq^{\pm2}-w)\over(z-w^{\pm2})} x^\pm(w)x^\pm(z)
\label{21d} \\
{[}x^+(z),x^-(w){]}&=& {(zw)^{-1}\over q-q^{-1}}\left(
\psi(wq^{1/2})\delta(z/qw)- \phi(zq^{1/2})\delta(w/qz)
\right)
\label{21e}
\eea
where $\delta(z)=\sum_{n\in/\ZZ}z^{n}$ is the delta function.

If we introduce the  operators $K$,
$a_n$, $a_{-n}$, $n\in\ZZ_+$, $x^\pm_m$,
$m\in\ZZ$ by means of the formulas
\bea
\psi(z)&=&\sum_{n=0}^\infty\psi_nz^{-n}=
K\exp\left((q-q^{-1})\sum_{n=1}^\infty a_nz^{-n}\right),\nn\\
\phi(z)&=&\sum_{n=0}^\infty\phi_nz^{n}=
K^{-1}\exp\left(-(q-q^{-1})\sum_{n=1}^\infty a_{-n}z^{n}\right),\nn\\
x^{\pm}(z) &=&\sum_{n\in\ZZ}x_n^\pm z^{-n-1}
\nn
\eea
then we can deduce from (\ref{21a})--(\ref{21e}) the commutation relations
\bea
{[}a_n,a_m{]}&=&\delta_{n,-m}{[2n][n]\over n},
\quad [a_n,K]=0,\nn \\
{[}a_n,x^\pm_m{]}&=&\pm{[2n]\over n}q^{\mp|n|/2}x^\pm_{n+m},\quad
Kx^\pm_mK^{-1}=q^{\pm2}x^\pm_m,\label{q-bosons}\label{cm2}\\
x^\pm_{n+1}x^\pm_{m}-q^{\pm2}x^\pm_{m}x^\pm_{n+1}&=&
q^{\pm2}x^\pm_{n}x^\pm_{m+1}-x^\pm_{m+1}x^\pm_{n},\nn\\
{[}x_n^+,x_m^-{]}&=&{1\over q-q^{-1}}
(q^{(n-m)/2}\psi_{n+m}-q^{(m-n)/2}\phi_{n+m}) \nn
\eea

We see that operators $a_n$ are free boson operators. Using commutation
relations (\ref{q-bosons}) one can find the bosonized expression for the
currents $x^\pm(z)$ \cite{FJ88}. But we would like to do the different
things. Namely, we want to show that
the new realization of the level one quantum
affine algebra $\Uqa$
follows from the free field representation
 of the ZF algebras (\ref{typeI}) and (\ref{typeII}).
As a by-product of this approach we  obtain
formulas for the Frenkel-Ding isomorphism between different
realizations of $\Uqa$.

\section{Bosonization of Zamolodchikov-Faddeev Algebras}

Let us start with bosonization of algebras (\ref{typeI}) and
(\ref{typeII}). First
note that due to (\ref{R-mat}) commutation relations for the operators
$\Phi_{\ep}(z_1)$, $\Phi_{\ep}(z_2)$  and $\Psi_{\nu}(z_1)$,
$\Psi_{\nu}(z_2)$ with the same isotopic indexes $\ep$, $\nu$
are simple
\bea
\Phi_\ep(z_2)\Phi_\ep(z_1)&=&r^+(z_1/z_2) \Phi_\ep(z_1)\Phi_\ep(z_2)
\label{32I}\\
\Psi_\nu(z_1)\Psi_\nu(z_2)&=&r^-(z_1/z_2) \Psi_\nu(z_2)\Psi_\nu(z_1).
\label{32II}
\eea
In order to solve (\ref{32I}) and (\ref{32II}) let us introduce
infinite dimensional Heisenberg algebra
$$
[a_n,a_m]=\delta_{n+m,0}s_n,\quad [\da,\alpha]=s
$$
where $s_n$ and $s$ are complex numbers and we would like to consider
them as parameters.  Also define the Fock space ${\cal F}$, where
the operators $\Phi_\ep(z)$ and $\Psi_\nu(z)$ act
\begin{equation}
{\cal F}=\hbox{linear span}\left\{
\prod_{j_k\geq \cdots\geq j_1>0}a_{-j_k}\cdots a_{-j_1}\right\}
\otimes e^{x\alpha}, \quad x\in\CC.
\label{Fock}
\end{equation}
In accordance with this definition of the Fock space, we understand
the normal ordered operator as an operator where all the negative  modes
 $a_{-n}$ and $\alpha$ are put on the left hand side of the positive
modes $a_n$ and $\da$.

Fix somehow $\ep$, $\nu$ and define the vertex operators
\bea
\Phi_{\ep}(z) =
\exp\left(\sum_{n=1}^{\infty}c_{-n}a_{-n}z^n\right)
\exp\left(\sum_{n=1}^{\infty}c_{n}a_{n}z^{-n}\right)
e^{c\alpha}z^{c'\partial_\alpha}
\label{33I}  \\
\Psi_{\nu}(z) =
\exp\left(\sum_{n=1}^{\infty}d_{-n}a_{-n}z^n\right)
\exp\left(\sum_{n=1}^{\infty}d_{n}a_{n}z^{-n}\right)
e^{d\alpha}z^{d'\partial_\alpha}
\label{33II}
\eea
where $c_{\pm n}$, $d_{\pm n}$, $c$, $c'$, $d$, $d'$
are parameters. Let us note first that normalization factors in
(\ref{32I}) and (\ref{32II}) can be written as follows
\bea
r^+(z)&=&\exp\left(-\sum_{k=1}^{\infty} {q^k\over k}{[k]\over[2k] }
(z^k-z^{-k})\right) z^{-1/2},\label{normI}\\
r^-(z)&=&\exp\left(\sum_{k=1}^{\infty} {q^{-k}\over k}{[k]\over[2k] }
(z^k-z^{-k})\right) z^{1/2}.\label{normII}
\eea
In order to obtain (\ref{normI}) and (\ref{normII}) we used the formula
\begin{equation}
(z)_\infty=\prod_{n=0}^{\infty}(1-zq^{4n})=\exp
\left(-\sum_{k=1}^{\infty}{1\over k}{z^k\over 1-q^{4k}}\right).
\label{34}
\end{equation}

On the other hand, the commutation relation for operators
(\ref{33I}) and (\ref{33II}) are
\bea
\Phi_\ep(z_2)\Phi_\ep(z_1)&=&
\exp\left(\sum_{n=1}^\infty s_nc_nc_{-n}
(z^n-z^{-n})\right)z^{-scc'}
\Phi_\ep(z_1)\Phi_\ep(z_2),
\label{332I}\\
\Psi_\nu(z_1)\Psi_\nu(z_2)&=&
\exp\left(\sum_{n=1}^\infty s_nd_nd_{-n}
(z^n-z^{-n})\right)z^{sdd'}
\Psi_\nu(z_2)\Psi_\nu(z_1).
\label{332II}
\eea
Identifying logarithms of the normalization factors
$r^\pm(z)$ and of those arisen in (\ref{332I}) and (\ref{332II}),
we obtain the following relations between parameters
\bea
s_nc_nc_{-n}&=&-{q^n\over n}{[n]\over[2n]},\quad scc'={1\over2},
\label{scc-rel}\\
s_nd_nd_{-n}&=&-{q^{-n}\over n}{[n]\over[2n]},\quad sdd'={1\over2}.
\label{sdd-rel}
\eea

It is clear now that we cannot use the same representation
(\ref{33I}) and (\ref{33II}) for the
operators $\Phi_{-\ep}(z)$ and $\Psi_{-\nu}(z)$
because in this case the relations
\bea
\Phi_{-\ep}(z_2)\Phi_{\ep}(z_1)=r^+(z_1/z_2)\left(
c(z_1/z_2) \Phi_{-\ep}(z_1)\Phi_\ep(z_2) +
b(z_1/z_2) \Phi_\ep(z_1)\Phi_{-\ep}(z_2) \right),
\label{middle1I}\\
\Phi_\ep(z_2)\Phi_{-\ep}(z_1)=r^+(z_1/z_2)\left(
\bar c(z_1/z_2) \Phi_{\ep}(z_1)\Phi_{-\ep}(z_2) +
b(z_1/z_2) \Phi_{-\ep}(z_1)\Phi_\ep(z_2) \right)
\label{middle2I}
\eea
and
\bea
\Psi_{-\nu}(z_1)\Psi_{\nu}(z_2)=r^-(z_1/z_2)\left(
c(z_1/z_2) \Psi_{-\nu}(z_2)\Psi_\nu(z_1) +
b(z_1/z_2) \Psi_\nu(z_2)\Psi_{-\nu}(z_1) \right) ,
\label{middle1II}\\
\Psi_\nu(z_1)\Psi_{-\nu}(z_2)=r^-(z_1/z_2)\left(
\bar c(z_1/z_2) \Psi_{\nu}(z_2)\Psi_{-\nu}(z_1) +
b(z_1/z_2) \Psi_{-\nu}(z_2)\Psi_\nu(z_1) \right).
\label{middle2II}
\eea
are not satisfied.
But it is still obvious that operators   $\Phi_{-\ep}(z)$  and
$\Psi_{-\nu}(z)$   should
be somehow proportional to the operator $\Phi_\ep(z)$ and
$\Psi_{\nu}(z)$  respectively in order to
cancel the common normalization factors $r^+(z_1/z_2)$ and $r^-(z_1/z_2)$.
To this end we introduce two more vertex operators
\bea
X(w) &=&
\exp\left(\sum_{n=1}^{\infty}x_{-n}a_{-n}w^n\right)
\exp\left(\sum_{n=1}^{\infty}x_{n}a_{n}w^{-n}\right)
e^{x\alpha}w^{x'\partial_\alpha}
\nn\\
Y(w) &=&
\exp\left(\sum_{n=1}^{\infty}y_{-n}a_{-n}w^n\right)
\exp\left(\sum_{n=1}^{\infty}y_{n}a_{n}w^{-n}\right)
e^{y\alpha}w^{y'\partial_\alpha}
\nn
\eea
and try to satisfy the commutation relations (\ref{middle1I}),
(\ref{middle2I}) and
(\ref{middle1II}), (\ref{middle2II}) by the linear combination
\bea
\Phi_{-\ep}(z)=(\Phi_{-\ep}(z,w))_m,\quad
\Phi_{-\ep}(z,w)=\Phi_\ep(z)X(w)+AX(w)\Phi_\ep(z),
\label{VOanzI}\\
\Psi_{-\nu}(z)=(\Psi_{-\nu}(z,w))_m,\quad
\Psi_{-\nu}(z,w)=\Phi_\nu(z)Y(w)+BY(w)\Psi_\nu(z),
\label{VOanzII}
\eea
where $A$, $B$ are parameter to be determined and notation $(O(w))_m$ means
$O_m$ if $O(w)=\sum_{m\in\ZZ}O_mw^{-m-1}$.

Some comment is in order now. Throughout this section we adopt the
``formal series'' point of view. This means that two formal
series (in spectral parameters) are equal if and only if all their
pairwise coefficients coincide.

Let us concentrate now on solving (\ref{middle1I}) and
(\ref{middle2I}). Relations (\ref{middle1II}) and
(\ref{middle2II})  can be treated analogously.  In order
to simplify the notations, let us introduce the new set of parameters
$$
\alpha_{-n}=-{x_{-n}\over c_{-n}}{q^n[n]\over[2n]},\quad
\alpha_{n}=-{x_{n}\over c_{n}}{q^n[n]\over[2n]},\quad
a={x\over2c},\quad a'= {x'\over2c'}
$$
such that the normal ordering products of the operators
$\Phi_{\ep}(z)$ and $X(w)$ can be written as follows
\bea
\Phi_{\ep}(z)X(w)&=&\exp\left(\sum_{n=1}^{\infty}
{\alpha_{-n}\over n}\left({w\over z}\right)^n\right) z^a
{:}\Phi_{\ep}(z)X(w){:}\ ,\label{ord1I}\\
X(w)\Phi_\ep(z)&=&\exp\left(\sum_{n=1}^{\infty}
{\alpha_{n}\over n}\left({z\over w}\right)^n\right) w^{a'}
{:}X(w)\Phi_\ep(z){:}\ .\label{ord2I}
\eea

Our strategy now is to find the relation between parameters
$\alpha_{\pm n}$
which fulfils (\ref{middle1I}), (\ref{middle2I})
 and
\begin{equation}
\Phi_{-\ep}(z_2)\Phi_{-\ep}(z_1)=
r^+(z_1/z_2) \Phi_{-\ep}(z_1)\Phi_{-\ep}(z_2) .
\label{lastI}
\end{equation}
First note that the zero mode part of operators
$\Phi_\ep(z)$ and $\Phi_{-\ep}(z)$
will be equal $e^{c\alpha}$ and
$e^{(x+c)\alpha}$ respectively. It is natural to require that
these operators has opposite zero modes, namely, to require that
$x+c=-c$.
It fixes the parameter $a$
$$
 a=-1.
$$
Now consider the equation (\ref{middle1I}).
Substitute there the operators $\Phi_{-\ep}(z_1,w)$ and
$\Phi_{\ep}(z_2)$ and
normal order all the operator products using (\ref{ord1I}) and (\ref{ord2I}).
It can be easily seen that the nontrivial equations
for the coefficients $\alpha_{\pm n}$ that follow from (\ref{middle1I})
can be obtained only if we set $a'=-1$.
This equations are
(the coefficient at $w^{-m-1}$ and $z^n$, $n,m\in\ZZ$ [$z=z_1/z_2$])
\bea
&
p_{n-m}(\alpha_-)p_n(\alpha_+)-q^2p_{n-m-1}(\alpha_-)p_{n-1}(\alpha_+)\nn\\
&\qquad+
A[p_{m-n-1}(\alpha_+)p_n(\alpha_+)-p_{m-n}(\alpha_+)p_{n-1}(\alpha_+)]\nn\\
&\quad=
(1-q^2-Aq)p_{-n}(\alpha_-)p_{m-n}(\alpha_+)+
Aqp_{-n-1}(\alpha_-)p_{m-n-1}(\alpha_+)\nn\\
&\qquad+
q[p_{-n-1}(\alpha_-)p_{n-m}(\alpha_-)-
p_{-n}(\alpha_-)p_{n-m-1}(\alpha_-)]  ,
\label{sub1}
\eea
where we introduced Schur polynomials
$p_{n}(\alpha_\pm)$ for the sets
$\alpha_{\pm}=(\alpha_{\pm1},\alpha_{\pm2}/2,\alpha_{\pm3}/3,\ldots)$
and put $p_{n}(\alpha_\pm)=0$ if $n<0$.
Analogous treatment of (\ref{middle2I}) yields the following relations
\bea
&
A[p_{n-m}(\alpha_-)p_n(\alpha_+)-q^2p_{n-m-1}(\alpha_-)p_{n-1}(\alpha_+)]\nn\\
&\qquad-
qA[p_{m-n-1}(\alpha_+)p_n(\alpha_+)-p_{m-n}(\alpha_+)p_{n-1}(\alpha_+)]\nn\\
&\quad=
-qp_{-n}(\alpha_-)p_{m-n}(\alpha_+)+
(A(1-q^2)+q)p_{-n-1}(\alpha_-)p_{m-n-1}(\alpha_+)\nn\\
&\qquad-
q^2[p_{-n-1}(\alpha_-)p_{n-m}(\alpha_-)-
p_{-n}(\alpha_-)p_{n-m-1}(\alpha_-)] .
\label{sub2}
\eea
Comparising  (\ref{sub1}) and (\ref{sub2}), one fixes $A=-q$ and, solving
these equations by induction, one gets the unique solution
\begin{equation}
\alpha_n=\alpha_{-n}=(-q)^n.
\label{solI}
\end{equation}

Some comment is in order now on ansatz (\ref{VOanzI}). It is clear
that we were sucsessful solving equations (\ref{sub1}) and (\ref{sub2})
only due to this ansatz. In fact, looking at these equations,
one can note that the other ansatz
\begin{equation}  \Phi_{-\ep}(z)=(\Phi_{-\ep}(z,w))_m,\quad
\Phi_{-\ep}(z,w)=z^{-1}(\Phi_\ep(z)X(w)+AX(w)\Phi_\ep(z))
\label{VOanzI2}
\end{equation}
leads to the same solution (\ref{solI}) but with $A=-q^{-1}$.
The existence of these simplest ansatz is due to the fact
that in the new realization of quantum affine algebra
$\Uqa$ there are four generators
that are proportional to corresponding Chevalley generators and
have simplest (two terms) action on the tensor products of two
representations of $\Uqa$.

With these solution for the parameters
$\alpha_\pm$ one can easily calculate the coefficients at
$(wv)^{-m}$ in the product $\Phi_{-\ep}(z_2,w)\Phi_{-\ep}(z_1,v)$
in order to obtain (up to function that arose from normal ordering of the
operators $\Phi_\ep(z_2)\Phi_\ep(z_1)$ and cancels the factor
$r^+(z_1/z_2)$)
$$
(-q)^{2m-1},\quad {\rm for}\quad m>0,\quad
(-q)^{-2m-1}(z_1z_2)^{m-1},\quad {\rm for}\quad m<0
\nn
$$
and
$$
q^{2}{[2]z_1z_2-q(z_1+z_2)^2\over z_1^2z_2^2},\quad {\rm for}\quad m=0.
$$
As far as these coefficients are symmetric functions of the
spectral parameters $z_1$ and $z_2$,
the equation (\ref{lastI}) is automatically satisfied
by solution (\ref{solI}) for any $m$ in (\ref{VOanzI}).

The analogous consideration of (\ref{middle1II}), (\ref{middle2II}) and
$$
\Psi_{-\nu}(z_2)\Psi_{-\nu}(z_1)=
r^-(z_1/z_2)\Psi_{-\nu}(z_1)\Psi_{-\nu}(z_2)
$$
yields the following unique solutions for the parameters
$$
\beta_{-n}=-{y_{-n}\over d_{-n}}{q^{-n}[n]\over[2n]}=(-q)^{-n},\quad
\beta_{n}=-{y_{n}\over d_{n}}{q^{-n}[n]\over[2n]}=(-q)^{-n},
$$
\begin{equation}
b={y\over2d}=-1,\quad b'= {y'\over2d'}=-1
\label{solII}
\end{equation}
with $B=-q$ in (\ref{VOanzII}) or with $B=-q^{-1}$ in  the ansatz
\begin{equation}
\Psi_{-\ep}(z)=(\Psi_{-\ep}(z,w)) _m,\quad
\Psi_{-\ep}(z,w)=z(\Psi_\ep(z)Y(w)+BY(w)\Psi_\ep(z)) .
\label{VOanzII2}
\end{equation}

Now we are in position to calculate  the operator product of the
operators $X(w)X(v)$ and $Y(w)Y(v)$
\bea
X(w)X(v)&=&\exp\left(-\sum_{n=1}^{\infty}(1+q^{-2n})
{\alpha_n\alpha_{-n}\over n} \left({v\over w}\right)^n\right)w^{-2a'}
{:}X(w)X(v){:}\nn\\
&=&(w-v)(w-q^2v){:}X(w)X(v){:}
\nn
\eea
\bea
Y(w)Y(v)&=&\exp\left(-\sum_{n=1}^{\infty}(1+q^{2n})
{\beta_n\beta_{-n}\over n} \left({v\over w}\right)\right)w^{-2b'}
{:}Y(w)Y(v){:}\nn\\
&=&(w-v)(w-q^{-2}v){:}Y(w)Y(v){:}
\nn
\eea
It follows  from these products that
the vertex operators $X(w)$ and
$Y(w)$ will satisfy the relations
\bea
X(v)X(w)&=& {(vq^{-2}-w)\over(v-w^{-2})} X(w)X(v)\nn \\
Y(v)Y(w)&=& {(vq^{2}-w)\over(v-w^{2})} Y(w)Y(v)\nn
\eea
that coincide with (\ref{21d}).

In order to calculate the commutation relations between operators $X(w)$ and
$Y(v)$ we have to use the relation (\ref{comm}). Comparising logarithm
of the  factor in the commutation relation
$$
\Psi_\ep(z_1)\Phi_\nu(z_2) =\exp
\left(\sum_{n=1}^{\infty}s_n(c_{-n}d_nz^{-n}-c_nd_{-n}z^n)\right)
z_1^{scd'}  z_2^{-sc'd} \Phi_\nu(z_2)   \Psi_\ep(z_1)
$$
with logarithm of the function $\tau(z)$
$$
\tau(z)=
\exp\left(-\sum_{k=1}^{\infty} {1\over k}{[k]\over[2k] }
(z^k-z^{-k})\right) z^{-1/2} ,\quad z=z_1/z_2
$$
gives the relations
\begin{equation}
s_n c_{-n}d_n  =s_n  c_nd_{-n}    ={1\over n}{[n]\over[2n]},\quad
scd'=sc'd=-{1\over2}.
\label{solIiII}
\end{equation}
These relations along with (\ref{scc-rel}), (\ref{sdd-rel}),
(\ref{solI}) and (\ref{solII}) allow one to express all parameters
only in terms of the parameters $c_{\pm n}$, $c$ and $c'$
$$
d_n=-c_nq^{-|n|},\quad x_n=-(-)^nc_n{[2n]\over[n]},\quad
y_n=(-q)^{-|n|}c_n{[2n]\over[n]},\quad n=\pm1,\pm2,\ldots
$$
\begin{equation}
-x=y=-2d=2c,\quad -x'=y'=-2d'=2c'
\label{paramet}
\end{equation}
It follows from (\ref{paramet}) that the operators
$$
[\Phi_\nu(z),Y(w)]=[\Psi_\ep(z),X(w)]=0
$$
commute, so the rest three relations in (\ref{comm}) are satisfied.

Commutation relation for operators $X(w)$ and $Y(v)$ can be easily
calculated now
$$
{[}Y(v),X(w){]}= {(vw)^{-1}\over q-q^{-1}}\left(
\psi(w)\delta(v/qw)
 -\phi(v)\delta(w/qv)
\right)
$$
where we introduced the operators
\bea
\psi(w)&=&{:}X(w)Y(qw){:}=
q^{2c'\da}\exp \left(\sum_{n=1}^{\infty}-(q-q^{-1})a_nw^{-n}c_n[2n]
(-q)^{-n}\right)
\nn\\
\phi(w)&=&{:}X(qw)Y(w){:}=
q^{-2c'\da}\exp \left(\sum_{n=1}^{\infty}-(q-q^{-1})a_{-n}w^{n}c_{-n}[2n]
(-)^{-n}\right)
 \nn
\eea
In addition, the simple calculation shows that
the commutation relations between operators
$\psi(z)$, $\phi(z)$  and
$\psi(z)$, $\phi(z)$, $X(w)$, $Y(w)$ are
\bea
\psi(z)\phi(w)&=&{(z-wq^3)(z-wq^{-3})\over(z-wq)(z-wq^{-1})}\phi(w)
\psi(z)\nn\\
\psi(z)Y(w)&=&q^{2}
{z-q^{-3}w\over z-q^{}w}Y(w)\psi(z) \nn\\
\psi(z)X(w)&=&q^{-2}
{z-q^{2}w\over z-q^{-2}w}X(w)\psi(z) \nn\\
\phi(z)Y(w)&=&q^{2}
{z-q^{-2}w\over z-q^{2}w}Y(w)\phi(z)   \nn\\
\phi(z)X(w)&=&q^{-2}
{z-q^{}w\over z-q^{-3}w}X(w)\phi(z)   \nn
\eea
and coincide  with  the
defining relation of $\Uqa$ in the new realization given by
(\ref{21a})--(\ref{21e}) up to some rescaling of the spectral parameters.

Let us summarize. From the bosonization of normalization factors
in commutation relations  of ZF algebras (\ref{typeI}),
(\ref{typeII}), (\ref{comm}) and from requirements that
operator equations (\ref{middle1I})--(\ref{middle2II}) to be
identified as formal power series with respect to all the spectral
parameters, one obtains the commutation relations of the level one
quantum affine algebra $\Uqa$ in the new realization.

Now we are ready to obtain formulas for the Frenkel-Ding isomorphism
between different realization of $\Uqa$.
Consider a following operator valued $2\times 2$ matrices
(indices $\nu$ and $\ep$ are fixed)
\bea
L^{(+)}(z,w,v)&=&
\left(\begin{array}{cc} 1&E^{(+)}(z) \\ 0&1  \end{array}\right)
\left(\begin{array}{cc} P^{(+)}(z)&0 \\ 0&K^{(+)}(z)  \end{array}\right)
\left(\begin{array}{cc} 1&0\\F^{(+)}(z) &1  \end{array}\right) \nn\\
&=&\sqrt z g  \left(
\begin{array}{cc}
\Psi_{-\nu}(z,w)\Phi_{-\ep}(z,v)&
\Psi_{-\nu}(z,w)\Phi_{\ep}(z) \\
\Psi_{\nu}(z)\Phi_{-\ep}(z,v) &
\Psi_{\nu}(z)\Phi_{\ep}(z)
\end{array}\right)
\nn
\eea
\bea
L^{(-)}(z,w,v)&=&
\left(\begin{array}{cc} 1&E^{(-)}(z) \\ 0&1  \end{array}\right)
\left(\begin{array}{cc} P^{(-)}(z)&0 \\ 0&K^{(-)}(z)  \end{array}\right)
\left(\begin{array}{cc} 1&0\\F^{(-)}(z) &1  \end{array}\right) \nn\\
&=&\sqrt z g  \left(
\begin{array}{cc}
\Phi_{-\ep}(z,w)\Psi_{-\nu}(qz,v)&
\Phi_{-\ep}(z,w)\Psi_{\nu}(qz) \\
\Phi_{\ep}(z)\Psi_{-\nu}(qz,v) &
\Phi_{\ep}(z)\Psi_{\nu}(qz)
\end{array}\right)
\nn
\eea
Simple calculations give the following relations
\bea K^{(+)}(z)
&=&q^{c'\da}\exp\left(\sum_{n=1}^{\infty}
c_{-n}(q^n-q^{-n})a_{-n}z^n\right)\nn\\
K^{(-)}(z)
&=&q^{-c'\da}\exp\left(\sum_{n=1}^{\infty}
q^{-n}c_{n}(q^n-q^{-n})a_{n}z^{-n}\right)\nn
\eea
Using commutativity of the operators $\Phi(z)$, $Y(w)$ and
$\Psi(z)$, $X(w)$, one can write
off-diagonal elements of $L$-operators as follows
\bea
E^{(+)}(z,w)K^{(+)}(z) &=&K^{(+)}(z)Y(w)-qY(w)K^{(+)}(z)
\label{Lp12}\\
K^{(+)}(z)F^{(+)}(z,v) &=&K^{(+)}(z)X(v)-qX(v)K^{(+)}(z)
\label{Lp21}\\
K^{(-)}(z)F^{(-)}(z,v) &=&K^{(-)}(z)X(v)-qX(v)K^{(-)}(z)
\label{Lm12}\\
E^{(-)}(z,w)K^{(-)}(z) &=&K^{(-)}(z)Y(w)-qY(w)K^{(-)}(z)
\label{Lm21}
\eea
Using the property of the delta function
$$
f(z)\delta(w/z)=f(w)\delta(w/z)
$$
and the variant of the formula (\ref{34})
$$
\exp\left(\sum_{n=1}^{\infty}{q^n-q^{-n}\over n}z^n\right)=
1+q^{-1}(q-q^{-1}) \sum_{n=1}^{\infty}  (qz)^n
$$
one can obtain from (\ref{Lp12})-(\ref{Lm21}) the Frenkel-Ding
formulas which relate the currents $X(w)$ and $Y(w)$ to elements of
the $L$-operators.
\bea
{E^{(+)}(-zq^{-1},w)-E^{(-)}(-zq^{-2},w)\over q-q^{-1}}&=&Y(z)\delta(z/w)
\nn\\
{F^{(-)}(-zq^{-1},w)-F^{(+)}(-zq^{-2},w)\over q-q^{-1}}&=&qX(z)\delta(z/w)
\nn
\eea
Using the commutation relations for the currents $X(z)$ and $Y(z)$
one can show that the ratio
$P^{(\pm)}(z,w,v)/K^{(\pm)}(z)$ is related to operators
$\phi(z)$ and $\psi(z)$                     as follows
\bea
P^{(+)}(-zq^{-1},w,v)(K^{(+)}(-zq^{-1}))^{-1}&=& -{1\over z^2}
\delta(v/qz)\delta(w/z)\phi(z),\label{PKphi}\\
P^{(-)}(-zq^{-1},w,v)(K^{(-)}(-zq^{-1}))^{-1}&=& -{1\over z^2}
\delta(v/z)\delta(w/qz)\psi(z).\label{PKpsi}
\eea

Until now we did not fix
the integer number $m$ in the ansatz for the operators
$(\!\Psi_{-\nu}(z,w))_m$ and $(\Phi_{-\ep}(z,w))_m$.
In order to fix it we have to use the relation which fixes
quantum determinants
of the operators $L^{(\pm)}(z)$. Let us calculate q-det$L^+(z,w,v)$.
Using obvious equality
$$
K^{(+)}(-zq^{-1}) K^{(+)}(-zq)=\phi(z)^{-1}
$$
and the commutation relation
$$
F^{(+)}(zq^{-2},v)K^{(+)}(z)=q K^{(+)}(z)  F^{(+)}(z,v)
$$
that follows from definition of $F^{(+)}(z,v) $
we obtain the relation
$$
\hbox{q-det} L^+(z,w,v)=-{1\over q^2z^2}\delta\left( -{v\over zq^{2}}\right)
 \delta\left( -{w\over zq}\right)
$$
Now it is clear that only the quantum determinant of the coefficient
at $(wv)^{-1}$ of the operator $L^+(z,w,v)$ satisfies (\ref{q-det}).
It means that in order to reconstruct a proper $L$ operators from
operators $\Psi(z)$ and $\Phi(z)$ using Miki's formulas (\ref{25})
we have to set $m=0$ in ansatz
(\ref{VOanzI}) and (\ref{VOanzII}). Similar arguments show that fixing the
quantum determinants along with ansatz
(\ref{VOanzI2}) for the operator $\Phi_{-\ep}(z,w)$
gives $m=1$ and doing this along with
ansatz (\ref{VOanzII2}) gives $m=-1$.

Now the operators
$\Phi_{\pm\ep}(z)$ and
$\Psi_{\pm\nu}(z)$ are completely determined and can be interpreted
as intertwiners of the corresponding highest weight
modules over quantum affine algebra $\Uqa$ at level one.
The Fock space (\ref{Fock}) decomposes into two subsppaces irreducible with
respect the action of $\Uqa$
$$
{\cal F}_i=\hbox{linear span}\left\{
\prod_{j_k\geq \cdots\geq j_1>0}a_{-j_k}\cdots a_{-j_1}\right\}
\otimes e^{(2n+i)c\alpha}, \quad n\in\ZZ,\quad i=0,1
$$
which can be identified with these modules.



\section{Conclusion}

To conclude, let us discuss possible extensions
of the approach developed in this paper.
First of all, this is  the case of higher levels for quantum affine
algebra $\Uqa$.
The commutation relations of ZF algebras are to be more
complicated \cite{IIJMNT92}, since they include the face $R$-matrices.
It is obvious that, in order to bosonize these ZF algebras,
one has to introduce, besides the free bosonic field, a $\beta\gamma$-system
similar to bosonization of arbitrary level algebra $\Uqa$ developed
in \cite{M94,Sh92}. In this case we meet the problem of decomposing
the huge Fock space to irreducible pieces that was solved recently
in \cite{K93} using some deformation of the Fedler construction.

Another possibile generalization
is to extend this analysis to $\Uqan$. In \cite{DO94}
the $R$-matrices which intertwine different finite dimensional
fundamental representations have been calculated and corresponding
ZF algebras have been defined. At level one, these algebras
were bosonized in \cite{Ko93}. The main problem here is to find a proper
generalization of Miki formulas for the $L$-operators. It can be easily
seen that naive generalization of (\ref{25}) to the case of algebra $\Uqan$
leads to incorrect commutation relations between
type I and type II vertex operators.

But the most interesting new development of our approach
would be to construct a
free field representation of the elliptic generalization of the affine
algebra $\widehat{\frak{sl}}_2$, which has been recently
formulated in
\cite{FIJKMY941,FIJKMY942} within the $R$-matrix approach.
Bosonization of the corresponding ZF algebras should allow
one to calculate the correlation functions and the form factors in the XYZ
model.

\section{Acknowledgments}

This work started during author's visit to RIMS. I would like
to acknowledge the members of this center for
a very stimulating scientific atmosphere.
Discussions with A.~Mironov, V.~Tarasov, F.~Smirnov and
J.~Schnittger were very helpfull.
The author would also like to acknowledge  Departamento de F\'\i sica
Te\'orica of Universidad de Zaragoza where this work has been completed.
This work   was supported by Direcci\'on General de
Investigaci\'on Cient\'\i fica y T\'ecnica (Madrid).



\end{document}